\documentclass[pre,showpacs,showkeys,preprintnumbers,amsmath,amssymb,superscriptaddress,onecolumn]{revtex4-1}
\usepackage{graphicx}
\usepackage{amsfonts}
\usepackage{color}
\usepackage{hyperref}

\def\<{\langle}
\def\>{\rangle}

\newcommand{\textgx}[1]{\textcolor{black}{#1}}

\begin{document}

\title{Concentrated suspensions of Brownian beads in water: dynamic heterogeneities trough a simple experimental technique}

\author{Raffaele Pastore}
\email{raffaele.pastore@unina.it}
\affiliation{Department of Chemical, Materials and Production Engineering,
University of Naples Federico II, P.le Tecchio 80, Napoli 80125, Italy.}

\author{Marco Caggioni}
\affiliation{Corporate Engineering, The Procter \& Gamble Company, Cincinnati, 8256 Union Centre Blvd., West Chester, OH 45069, USA}

\author{Domenico Larobina}
\affiliation{Institute for Polymers, Composites and Biomaterials, National Research Council of Italy, P.le E. Fermi 1, Napoli, 80055 Portici, Italy}

\author{Luigi Santamaria Amato}
\affiliation{ASI, Agenzia Spaziale Italiana, Centro di Geodesia Spaziale, 75100 Matera , Italy}

\author{Francesco Greco}
\affiliation{Department of Chemical, Materials and Production Engineering,
University of Naples Federico II, P.le Tecchio 80, Napoli 80125, Italy.}

\date{Received: \today / Revised version: }

\begin{abstract}
Concentrated suspensions of Brownian hard-spheres in water are an epitome for understanding 
the glassy dynamics of both soft materials and supercooled molecular liquids. 
From an experimental point of view, such systems are especially suited to perform particle tracking easily, and, 
therefore, are a benchmark for novel optical techniques, applicable when primary particles cannot be resolved.
Differential Variance Analysis (DVA) is one such novel technique that
simplifies significantly the characterization of structural relaxation processes of soft glassy materials, since it is
directly applicable to digital image sequences of the sample.
DVA succeeds in monitoring not only the average dynamics, but also its spatio-temporal fluctuations, known as dynamic heterogeneities.
In this work, we study the dynamics of dense suspensions of Brownian beads in water, imaged through digital video-microscopy, 
by using both DVA and single-particle tracking. 
We focus on two commonly used signatures of dynamic heterogeneities:  
the dynamic susceptibility, $\chi_4$, and the non-Gaussian parameter, $\alpha_2$.
By direct comparison of these two quantities, we are able to highlight similarities and differences.
We do confirm that $\chi_4$ and $\alpha_2$ provide qualitatively similar information, but we find
quantitative discrepancies in the scalings of characteristic time and length scale on approaching the glass transition.

\end{abstract}



\maketitle
\section{Introduction}
The structural relaxation of soft materials, such as polymer melts, colloidal suspensions and emulsions,
 plays a crucial role in tailoring their mechanical and rheological properties.
In particular, a dramatic increase of relaxation times accompanied by growing dynamic heterogeneities is the hallmark of soft materials 
approaching a structural arrest transition~\cite{DHbook, BerthierPhys}.
Dense suspensions of Brownian hard-spheres in water were a 
pioneeristic model system~\cite{pusey1986phase} to experimentally 
validate predictions for dynamic heterogeneities~\cite{WeeksScience,kegel2000direct}, 
first coming from theoretical models and numerical simulations~\cite{glotzer2000spatially,bennemann1999growing}.
This system is still today very popular to test novel features of arrested materials and 
to challenge unsolved issues on the 
glass transition~\cite{nagamanasa2015direct,gokhale2016localized,ganapathi2018measurements,hallett2018local,robinson2018morphometric}. 
The dynamic heterogeneity scenario emerging from hard-sphere suspensions has been then confirmed in a wide variety of systems, 
including  different types of colloidal glasses~\cite{CipellettiNatPhys,colin2015questioning}, gels~\cite{Trappe,philippe2017mucus}, granular materials~\cite{Keys},
cell tissues~\cite{Fredberg}, and many complex formulated products~\cite{Spicer}.
Quite recently, dynamic heterogeneities have been also measured in supercooled water~\cite{perakis2017diffusive}. 
These measurements require tremendous efforts, because of the atomistic nature of the primary particles in molecular liquids
and, in fact, have become available only in the last few years.   

In general, the experimental characterization of dynamic heterogeneities still remains a complicated task even for
non-molecular liquids, and is indeed handled only by specialized groups, 
since it is required to resolve the dynamics in space and time, and to estimate deviations from the
average behavior.
Dynamic Light Scattering techniques are probably the most robust quantitative methods~\cite{CipellettiNatPhys,DWS},
and allow for measuring the most popular estimator of dynamic heterogeneities, namely, the dynamic susceptibility, $\chi_4$. 
However, Dynamic Light Scattering does not provide any direct visualization of dynamic heterogeneities. 
Simultaneous visualization and quantitative measurements require more 
sophisticated techniques, including approaches that combine features of both Dynamic Light Scattering and imaging~\cite{CipellettiPRL09}. 
In flowing granular materials or jammed colloidal suspensions, when thermal motion is negligible, 
dynamic heterogeneities have been monitored using simpler methods, based on autocorrelation of image intensity (ACII) ~\cite{Durian_SM, Durian_PRE}.
Differential Dynamic Microscopy~\cite{DDM,Cerbino_Cicuta} is an elegant and promising technique, 
but currently limited to monitor the structural relaxation, and not dynamic heterogeneities.
Single particle tracking can be, of course, exploited when particles are clearly resolved by the microscope~\cite{WeeksScience}  but,
even in this case, quantitative measurements of $\chi_4$ suffer from major limitations and may be not statistically meaningful, as discussed in Sec.~\ref{Subsec:ngp} .
Nevertheless, other dynamic quantities can be readily computed via particle tracking,
including different types of dynamic correlation functions 
and indirect dynamic heterogeneity indicators, such as Van Hove functions and non-Gaussian parameters~\cite{WeeksScience,kegel2000direct}. 
Particle tracking also represent a benchmark in developing fully optical techniques. 
Indeed, it is fundamental to interpret the bulk observables as obtained from the latter, and to test their precision.  
Even in this technical perspective, a system of Brownian hard beads in water is ideal to easily perform single particle tracking,
also in crowded conditions, due, for example, to the good contrast typically present between particle and water and
to well-defined particle shape.  

Since soft glassy materials are widespread in industry and biology,
an easy optical method to characterize the relaxation of complex fluids with dynamic heterogeneity appears to be really on demand.
Some of us recently introduced Differential Variance Analysis (DVA)~\cite{DVA}, a method  
that can be directly applied to digital videos of a sample, without any need of tracking single particle positions.
DVA focuses on the {\it differential frames}, obtained by subtracting images at different lag-times;
a dynamic order parameter for structural relaxation and the dynamic susceptibility are simply obtained 
from real space variance of differential frames and its fluctuations, respectively.
Direct visualization of dynamic heterogeneities is quickly achieved, by looking at differential frames at lag-times around the
maximum dynamic susceptibility.

In this paper, we draw on video microscopy of colloidal suspensions of silica hard spheres in water,
 with the dynamics slowing down on increasing the volume fraction.
After  an overview of the DVA workflow, we illustrate by direct visualization how the shape of dynamic heterogeneities 
evolves in time, and on varying the volume fraction. 
Performing particle tracking, we then compute the lowest-order non-Gaussian parameter $\alpha_2$, 
a commonly used, yet indirect, indicator of dynamic heterogeneities(see for example Ref.s~\cite{WeeksScience,kegel2000direct,Chi_Gnan,Baschnagel2018} . 
We can therefore compare the DVA-based dynamic susceptibility $\chi_4$ to $\alpha_2$.
Those quantities are expected to give similar information on dynamic heterogeneities,
although a direct comparison is not usually performed.
We do find that, in fact, $\chi_4$ and $\alpha_2$ display an analogous behaviour; 
however, some quantitative differences in the scalings of characteristic time and length scale on approaching the glass transition do also show up.

\section{Materials and Method}

\label{Sec:methods}
We utilize data from previous experiments~\cite{SM15_exp}, which investigated the intermittent single particle motion using particle tracking. 
Quasi-two dimensional hard-sphere-like colloidal suspensions at different volume fractions, $\Phi$, were obtained
by using a 50:50 binary mixture of silica beads dispersed in a water.  Surfactant (Triton X-100, 0.2\% v/v) was added to the solution
to avoid particle sticking. 
Large and small beads diameters measure $d_l=3.16$ and $d_s=2.31$ $\mu m$, respectively, resulting in an $1.4$, ratio known to prevent crystallization.
Digital videos of the samples were obtained using a standard microscope equipped with a 40x objective (OlympusUPLAPO 40XS) and 
a fast digital camera (Prosilica GE680). 
We focused on a volume fraction range where the samples can be equilibrated on the experimental timescale, and monitored
the dynamics after equilibrium is attained. 
At the highest volume fraction, 
roughly a thousand particles in the field of view of the microscope were imaged.
At each volume fraction, the video duration, $t_v$, was several times larger than the relaxation time, $\tau$, 
while the interval between subsequent frames, $t_f$, was much smaller than $\tau$.
In particular, $t_v$ ranged in $[10^3 s, 10^5 s]$ and $t_f$ in $[0.2s, 2 s]$, respectively, 
depending on the volume fraction, i.e., larger times at larger volume fraction.
(See Appendix A1 for a discussion on general video requirement to properly perform DVA.)
Data analysis was performed using Python and SciPy libraries~\cite{Python1}.
Interactive data exploration and visualization was performed using IPython and Jupyter notebooks~\cite{Python2}.
DVA code is freely available at the corresponding author web-page, \url{http://rpastore.altervista.org}.

\section{Results}
\subsection{DVA workflow}
\begin{figure}[h!]
\centering
\includegraphics[scale=0.5] {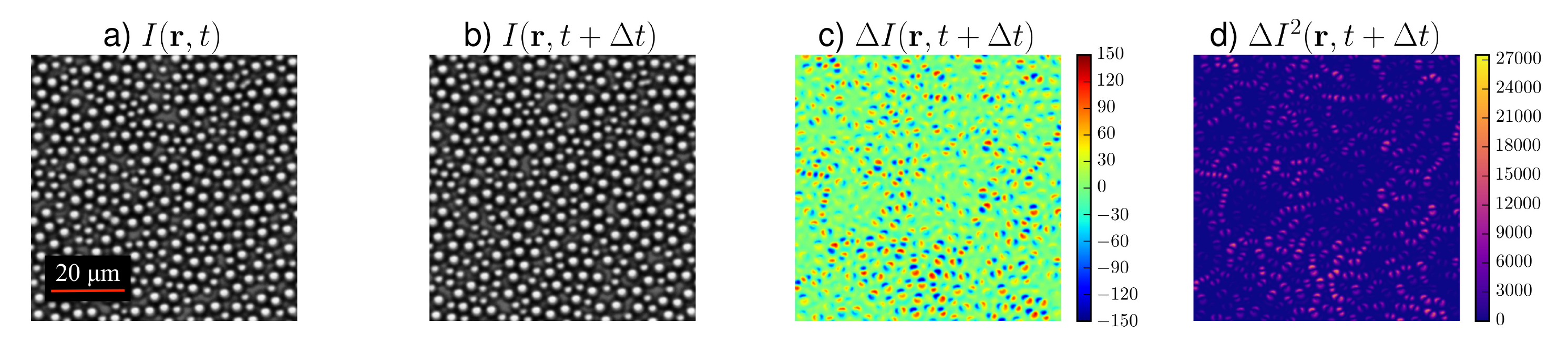} 
\includegraphics[scale=0.9]{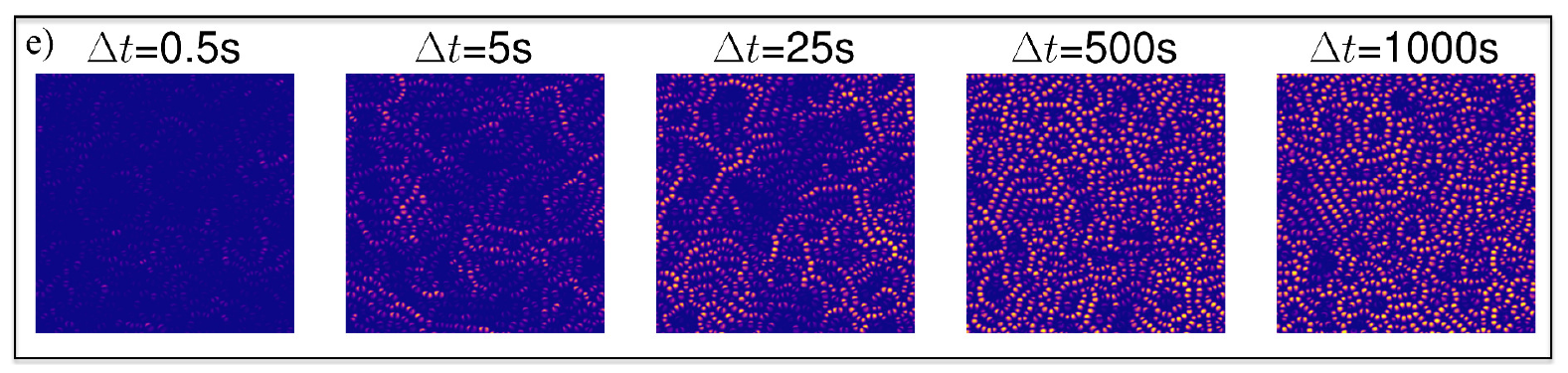}

\includegraphics[scale=0.3]{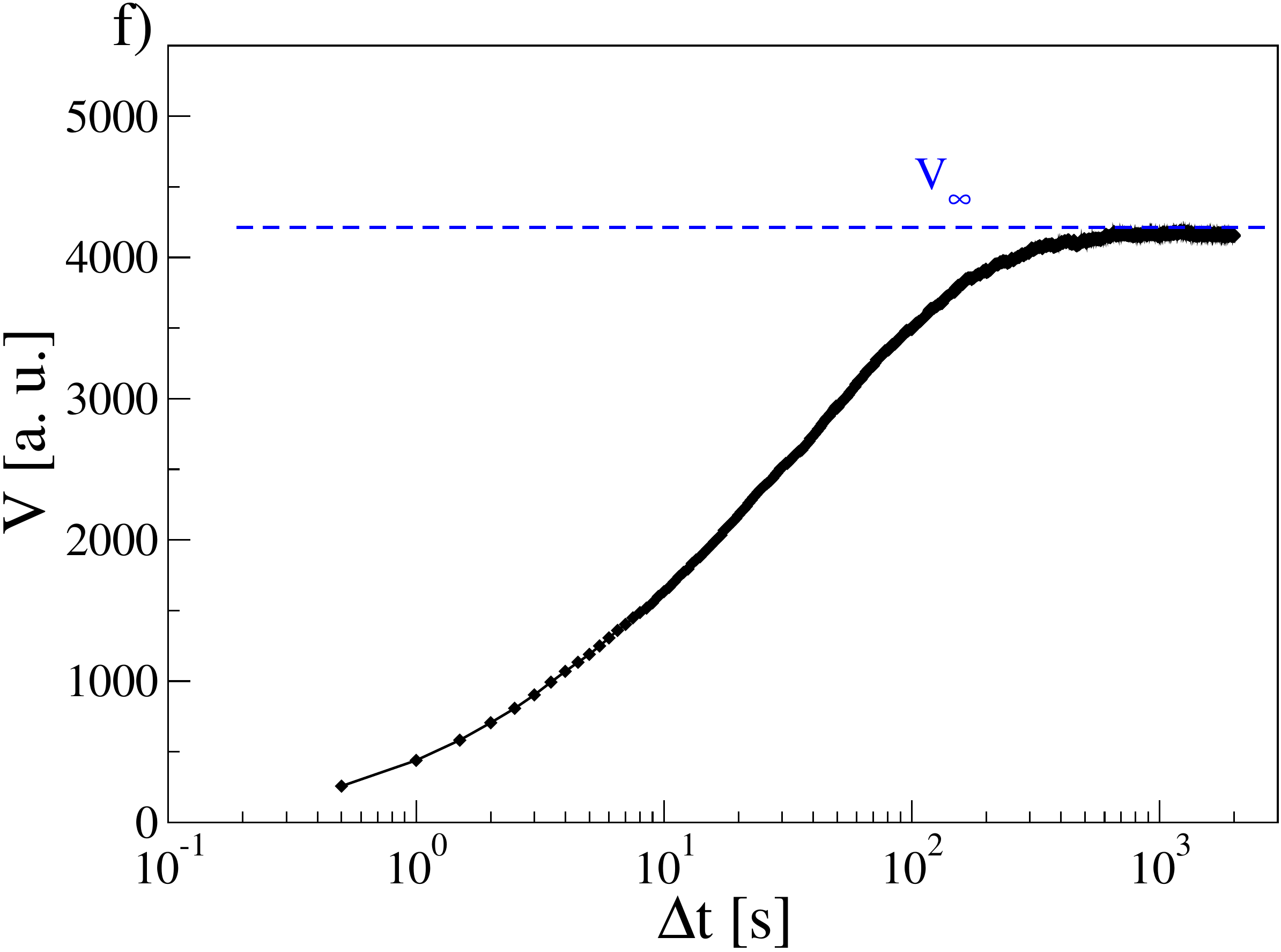}
\includegraphics[scale=0.3]{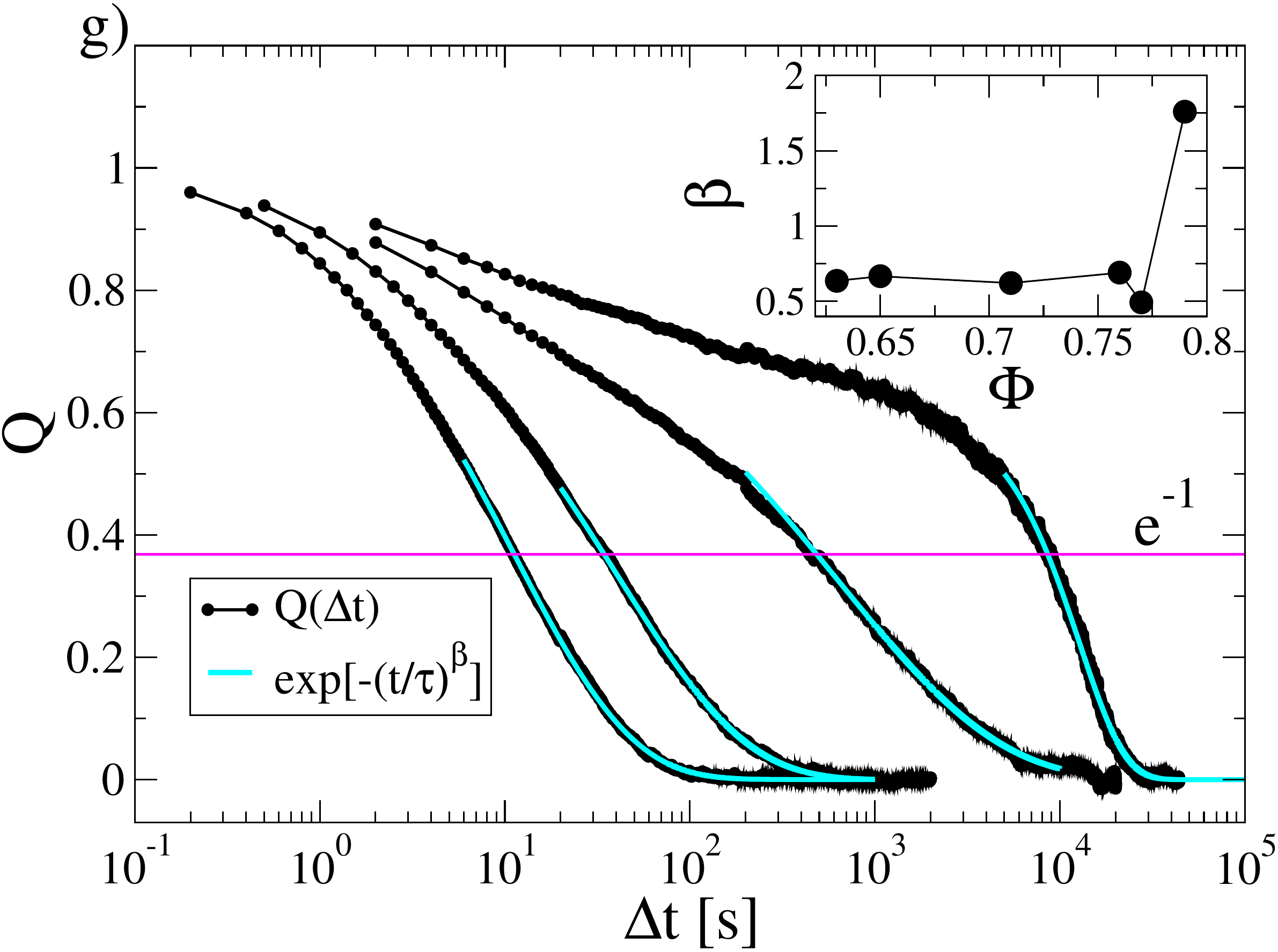}

\includegraphics[scale=0.3]{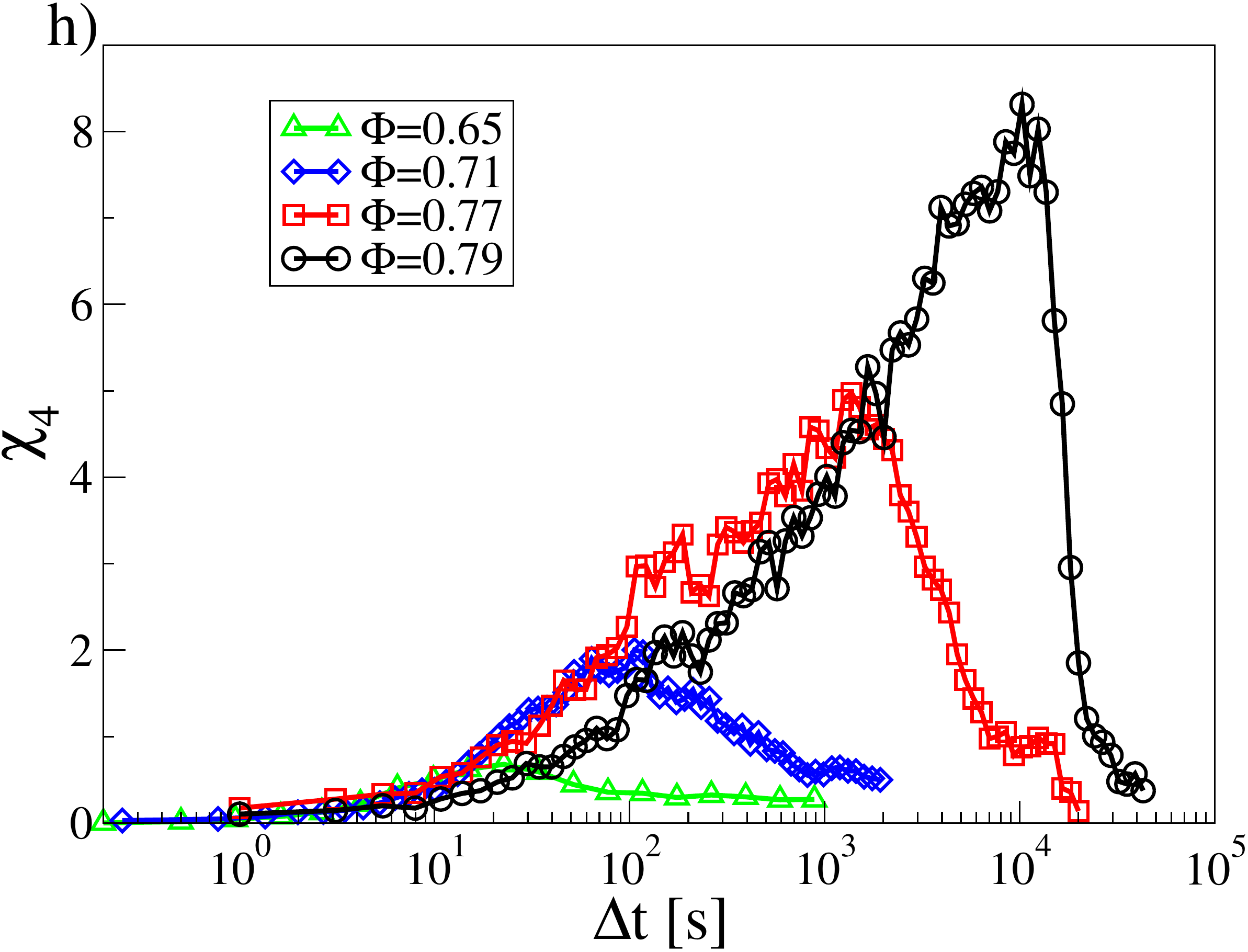}
\includegraphics[scale=0.3]{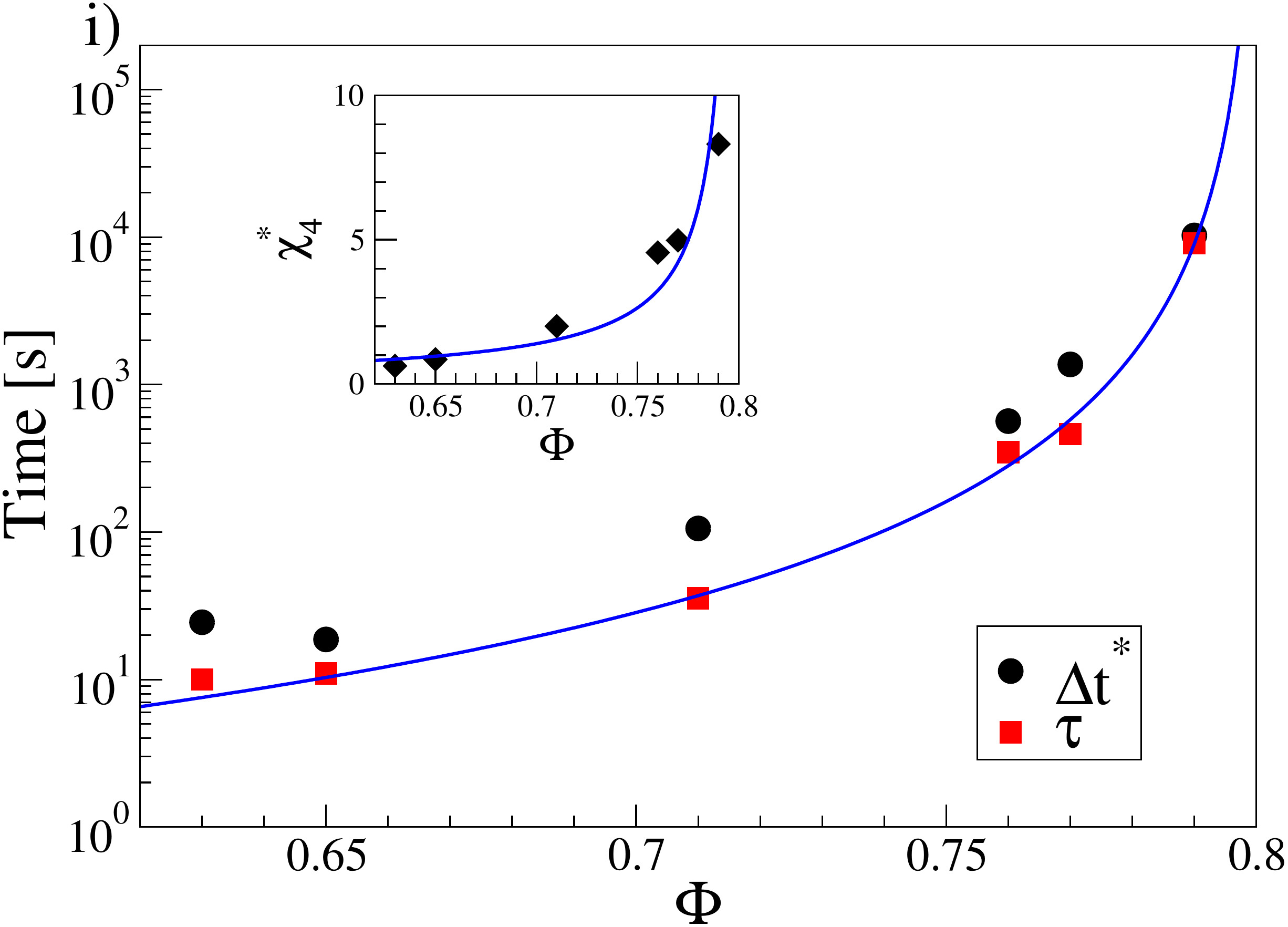}
\caption{\label{fig:fig1}
For  a sample at volume fraction, $\Phi=0.71$:  (a, b) two frames of a portion of a sample separated by a time-lag, $\Delta t=10 s$.  
(c) The resulting differential frame and (d) the square of this latter (color-scales as indicated). 
(e) Sequence of square differential frames at different times-lags, $\Delta t$, as indicated (color-scale as in panel (d)). 
(f) Average intensity variance of differential frames, $V$, as a function of the lag-time $\Delta t$. 
The dashed line indicates the long time plateau $V_{\infty}$.
(g) For samples at different volume fractions, $\Phi=0.65, 0.71, 0.77, 0.79$ (from left to right):
 DVA dynamic order parameter, $Q(\Delta t)$. Solid lines are Kohlrausch-Williams-Watts fit to the late decay.
(h) Dynamic susceptibility, $\chi_4$, as function of the time-lag $\Delta t$:
(i) Relaxation time, $\tau$, and the time corresponding the maximum susceptibility, $\Delta t^*$, as a function of $\Phi$.  
$\tau$ is fitted by a power-law $\propto (\Phi_c-\Phi)^{-\gamma}$ (solid line), with $\gamma=2.5\pm0.2$.
Inset: Maximum dynamic susceptibility, $\chi_4^*$,  as a function of the volume fraction, $\Phi$.  The solid line is a power-law 
$ \propto (\Phi_c-\Phi)^{-\alpha}$, with $\alpha=0.9\pm0.1$.
Adapted from Ref.~\cite{DVA} under a Creative Commons Attribution 4.0
International License.  
}
\end{figure}

The first step in the DVA approach consists in considering, from  a digital video of a sample, couples of frames characterized by a lag-time $\Delta t$, and,
for each couple,
the \textit{differential frame} obtained by subtraction of the pixel intensities of the two frames, $\Delta I({\bf r}, t,\Delta t)=I({\bf r},t+\Delta t)-I({\bf r},t)$. 
As an illustrative example, Figure~\ref{fig:fig1} shows in panels a and b two frames separated by a lag-time $\Delta t=10 s$ for a system of volume fraction  
$\Phi=0.71$ and, in panel c, the resulting differential frame.  
The lag-time of panel c is slightly smaller of the structural relaxation time of this system. 
On this timescale, some particles are expected to stay still localized close to their initial position (slow particles), while
other particles have moved over a distance comparable to their size (fast particles)~\cite{WeeksScience}.
This separation clearly appears in the differential frame:
i) the green background corresponds to the slow particles,
which rattle around their original positions, resulting in small  deviations around $\Delta I = 0$;
ii) the fast particles give rise to close spots of 
negative (blue) and positive (red) $\Delta I$, which look like dipoles.
These dipoles not only signal the positions of the fast particles, but also provide information on the direction of the displacement, 
with the blue spot corresponding to the position occupied by the particle at time $t$ but not at time $t+\Delta t$, and vice-versa for a red spot. 
As far as one is not interested in the direction of the displacements, 
a similar but slightly simpler visualization can be obtained by considering the square of the differential frame,  $[\Delta I({\bf r}, t,\Delta t)]^2$, as in panel d. 
Figure~\ref{fig:fig1}e illustrates the system temporal evolution
through a sequence of squared differential frames at increasing time-lag.
This allows for a direct visualization of the structural relaxation process as time goes on, namely, 
how the initial configuration at time $t$ becomes progressively uncorrelated with that at time $t+\Delta t$, upon increasing $\Delta t$.
Initially, the number of spots in the frames of panel e increases as $\Delta t$ increases, as more and more particles have moved.
Conversely, in the long time limit, i.e. for lag-times much larger than the relaxation time,
all particles have moved away from their original position and the number of spots seems to saturate.
Notice that low-intensity spots survive in the long-time limit, but they are due 
to the random overlap of different particles at the initial and final time, rather than to the presence of "frozen" particles.

The space integral of the squared differential frame directly defines the variance of $\Delta I$,
\begin{equation}
\label{eq:V}
\hat V(t, \Delta t)=\frac{1}{L^2} \int_{L^2}  [\Delta I({\bf r},t,\Delta t)]^2 ~d{\bf r},
\end{equation}
with $L$ the size of the frame.
As the investigated systems are in equilibrium conditions, and, therefore, the dynamics is invariant under translation of the initial time, $t$,
we define the $t$-independent variance  of $\Delta I$, $V (\Delta t)=\< \hat V(t, \Delta t) \>_t$,  
obtained by averaging $\hat V(t, \Delta t)$ over the ensemble of differential frames with different initial times $t$
and the same $\Delta t$.
The variance $V (\Delta t)$, shown in Fig.~\ref{fig:fig1}f, fully reflects the dynamic behavior illustrated in Fig.~\ref{fig:fig1}e, 
which refers to the same volume fraction, $\Phi=0.71$.
Indeed, $V (\Delta t)$ increases as far as the number of spots in the squared differential frame increases,  
whereas  it approaches a constant value, $V_{\infty}$, at long times, when the number of spots saturates. 
Hence, the (t-independent) differential frame variance $V (\Delta t)$ can be exploited to describe the relaxation process.

Notice that, generally, structural relaxation
is described through dynamic correlations functions, 
which essentially account for the fraction of particles that, after a time $\Delta t$, have moved less than an assigned
probing length scale, $\lambda$: therefore,  dynamic correlations functions decay from 1 to 0 as the system relaxes.
In glassy materials, different measured probes are known to provide similar information~\cite{DHbook, BerthierPhys},
one of the most popular  dynamic correlations function being the Intermediate Self Scattering Function.
By choosing $\lambda$ of the order of particle size, dynamic correlations functions are termed {\it Dynamic order parameter}. 

To match this standard framework, we introduce a variance based Dynamic order parameter:
\begin{equation}
\label{eq:overlap}
\hat Q(t, \Delta t)=1-\frac{\hat V(t, \Delta t)}{V_{\infty}},
\end{equation}
and its (t-independent) average, $Q(\Delta t)=\< \hat Q(t, \Delta t) \>_t$, defined analogously to $V (\Delta t)$.
Figure~\ref{fig:fig1}g shows $Q(\Delta t)$ at different volume fraction,  
clarifying that the decay becomes slower and slower as the glass transition is approached, on increasing $\Phi$.
We demonstrated, in a previous paper~\cite{DVA}, that the differential frame variance can be used to  quantitatively describe the relaxation process:
indeed, we found that $Q(\Delta t)$ equals to very good approximation the Intermediate Self Scattering Function (measured by particle tracking)
at a wavelength of the order of the interparticle distance, for all the investigated volume fractions. 
(See Appendix A2 for a comparison between the Dynamic order parameters of DVA and ACII). 
Note that, on changing the volume fraction in the investigated range, i.e. $\Phi \in.[0.63, 0.79]$, a very small change of the interparticle distance results. 
Indeed, the average interparticle distance is proportional to $\Phi^{-1/3}$ and, therefore, the ratio between the interparticle distances at the smallest and the largest $\Phi$
 is $(0.63/0.79)^{-1/3}=1.08$. Thus, in the whole range of investigated volume fractions, the typical interparticle distance varies by $8\%$ only. 
The relaxation time $\tau$,  conversely, changes by three orders of magnitude. 
This is a typical signature of glassiness: dramatic dynamical changes are accompanied by minor structural changes. 

The simplest signature of the dynamic slowing down, which takes place on approaching the glass transition,
is indeed the structural relaxation time, $\tau$.
Once clarified that $Q(\Delta t)$ is an effective Dynamic order parameter of the structural relaxation process,
we can simply estimate this characteristic time using the relation $Q(\tau)=1/e$.
Figure~\ref{fig:fig1}i shows that $\tau$ grows on increasing the volume fraction, and is compatible with a power-law, $(\Phi_c-\Phi)^{-\gamma}$,
as predicted  by Mode Coupling Theory. In particular, we find  $\Phi_c\simeq0.80\pm0.01$ and $\gamma=2.5\pm0.2$.
Notice that the late decay of $Q(\Delta t)$ is well fitted by Kohlrausch-Williams-Watts law $Ae^{-({t/\tau})^{\beta}}$ (solid lines in Fig.~\ref{fig:fig1}g),
a common property of Dynamic order parameters in glass-formers~\cite{DHbook}. The exponent  stays around $\beta\simeq0.55$ 
for all but the last two volume factions investigated, to slightly decrease at $\Phi=0.77$ and suddenly jump to $\beta\simeq 1.75$ for the densest system (see Inset of Figure~\ref{fig:fig1}g). 
A similar change from stretched ($\beta<1$) to compressed exponential ($\beta>1$) 
has been previously reported in other glassy systems, 
including nearly hard sphere~\cite{CipellettiNatPhys}
and soft colloidal glasses~\cite{Angelini2018}, where a minimum in $\beta$ is also found. 
Different numerical models of amorphous solids and glass-formers have been recently proposed to capture this stretched-to-compressed 
exponential crossover~\cite{Delgado,Ferrero,Gnan2018,wu2018stretched},
but its general understanding is still under debate. 
We notice that, in our case, the sharp increase of $\beta$ at the largest volume fraction may signal a breakdown of
the Mode Coupling regime and the onset of a different regime:  it might be associated to the 
famous "fragile-to-strong" crossover observed in supercooled molecular liquids~\cite{Taborek1986,saika2001fragile}, 
especially 
water~\cite{Angell1999,Chen2006,Nilsson2009,Mallamace2009,Mallamace2010,Gallo2016,Stanley2007,Xu2009,Gallo_review2016}, 
and in colloidal glasses~\cite{Mallamace2013}, also as a consequence of aging ~\cite{Angelini2013,Angelini2015}.
Association of strecthed-to-compressed and fragile-to-strong crossovers can also be inferred from recent experiments on microgel and metallic glasses~\cite{Angelini2018}. 
 
The  relaxation time $\tau$ is, of course, indicative of the average dynamics.
However, this information poorly characterizes dynamic heterogeneous systems, such as glass--forming liquids,
since individual particle behaviour can strongly deviate from the average behaviour, at least up to timescales of the order of the relaxation time. 
Indeed, in liquids close to the glass transition, dynamic heterogeneities emerge as transient clusters of particles with a mobility larger or smaller than the average~\cite{WeeksScience}.
The size and the lifetime of these dynamical clusters increase on approaching the transition, 
playing a role similar to density fluctuations close to an ordinary critical point~\cite{Halperin,Whitelam,Fractals}.
According to a robust framework developed by the glass community~\cite{DHbook, BerthierPhys}, 
the degree of dynamic heterogeneity can be monitored trough the dynamic susceptibility, $\chi_4(\Delta t)$,
defined from Dynamic order parameter fluctuations.
Within DVA, dynamic susceptibility is therefore defined from the fluctuations of our Dynamic order parameter: 
\begin{equation}
\label{eq:chi4_3}
\chi_4(\Delta t)=N\left [ \< \hat Q^2(t, \Delta t) \>_t - \< \hat Q(t, \Delta t) \>_t^2 \right ],
\end{equation}
$N$ being the number of imaged particles, which can be estimated from the relation $N=\frac{8\Phi L^2}{\pi(d_s^2+d_l^2)}$.
In terms of $\hat V$, the dynamic susceptibility can be directly related to the variance,
$\chi_4(\Delta t)=\frac{N}{V_{\infty}^2} \left [\< \hat V^2(t, \Delta t) \>_t - \< \hat V(t, \Delta t)\>_t^2 \right ]$.

Figure~\ref{fig:fig1}h shows that $\chi_4(\Delta t)$ has a maximum $\chi_4^*$ at a time $\Delta t^*$,
and that both those quantities increase on increasing the volume fraction, as typically  found~\cite{DHbook}.
$\chi_4^*$ and $\Delta t^*$ are a rough estimate of the typical size and life-time, respectively, of a cluster of dynamically correlated particles.
Thus, these dynamical clusters 
become increasingly spatially extended and long-lived as glass transition is approached~\cite{DHbook},
resembling static density fluctuations in ordinary critical phenomena.

Fig.~\ref{fig:fig1}i shows quantitatively how  $\Delta t^*$ spans almost three orders of magnitude, 
mimicking the behaviour of the relaxation time, $\Delta t^* \simeq \tau$.
The inset of Fig.~\ref{fig:fig1}i clarifies that also $\chi_4^*$  may be compatible with a power-law in $(\Phi_c-\Phi)$.

\subsection{DVA-based Direct visualization}
\begin{figure}[ht!]
\centering
\includegraphics[width=\linewidth]{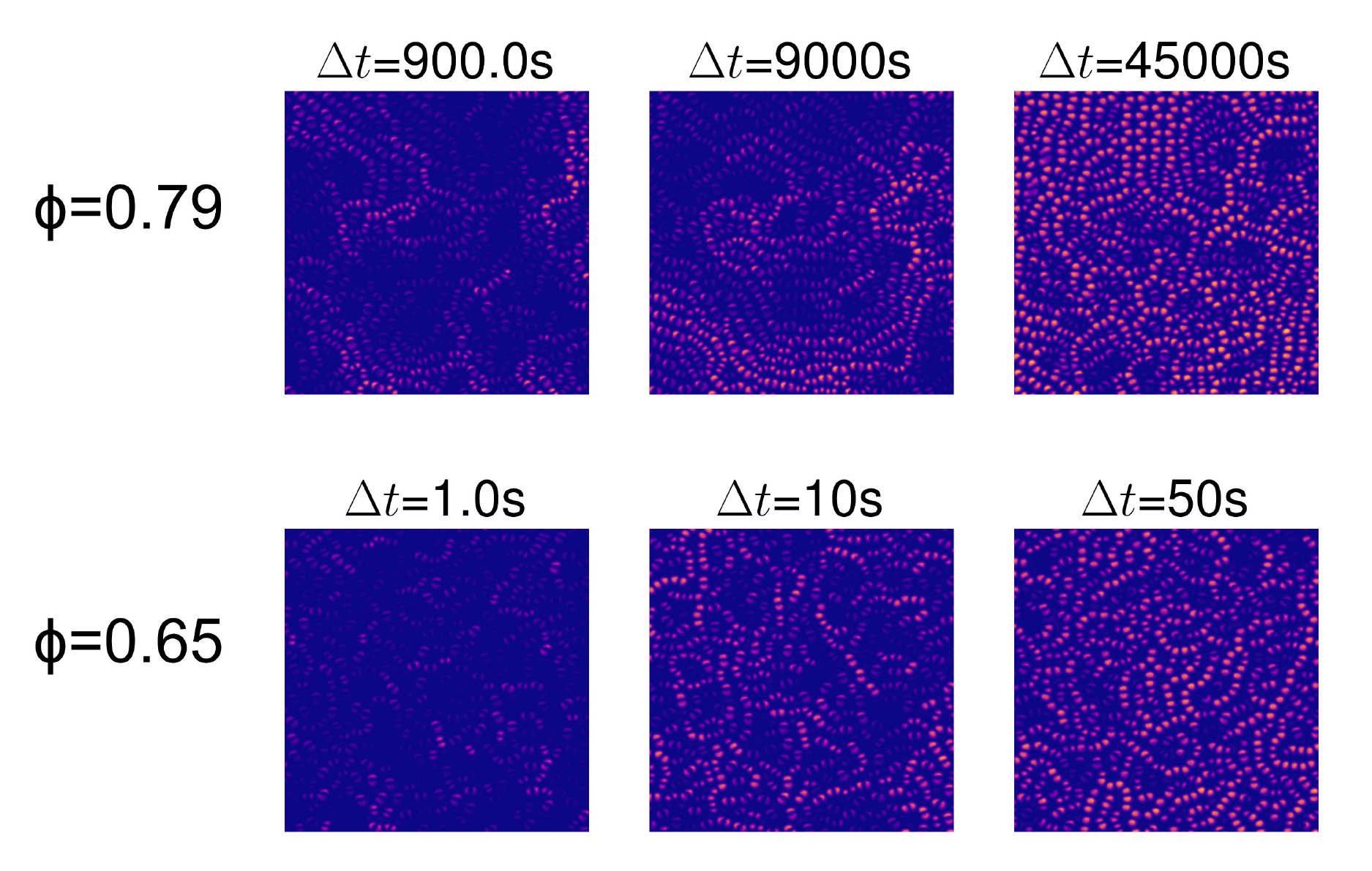}
\caption{\label{fig:direct} 
Two sequences of square differential frames at increasing $\Delta t$ at $\Phi=0.79$ (top) and $\Phi=0.65$ (bottom), as indicated.
From left to right, $\Delta t\simeq 10^{-1}\tau$, $\Delta t\simeq \tau \simeq \Delta t^*$ and $\Delta t\simeq 10\tau$,
with the relaxation time being $\tau \simeq 10^4 s$ and $\tau \simeq 10^1 s$ at $\Phi=0.79$ and $\Phi=0.65$, respectively.
} 
\end{figure}

$\chi_4$ provides quantitative information on the temporal evolution of dynamic heterogeneities and their dependence on the volume fraction.
Now we show that this quantitative information has a direct visual counterpart in the differential frames, if one focuses on the proper timescale.
To this aim, Fig.~\ref{fig:direct} shows two sequences of three differential frames for a large and a small value of the volume fraction, $\Phi=0.79$ and $\Phi=0.65$, respectively.
Moving along a row, the lag-time, $\Delta t$, is increasing and 
the system is progressively relaxing with respect to the initial configuration.
For each volume fraction, the lag-time are chosen so that $\Delta t\simeq 10^{-1}{\tau}$, $\Delta t\simeq \tau \simeq \Delta t^*$ and $\Delta t\simeq 10\tau$ (from left to right),
with the relaxation time being $\tau \simeq 10^4 s$ and $\tau \simeq 10 s$ for $\Phi=0.79$ and $\Phi=0.65$, respectively (see Fig.~\ref{fig:fig1}i).
From Fig.~\ref{fig:fig1}h, we learnt that at $\Delta t$ values significantly smaller or larger than $\tau$, $\chi_4$ is, in general, relatively small.
Accordingly, we see that the first and last 
differential frame of each sequence show poor correlations in the spatial distributions of spots.
In particular, spots are rare and sparse at short time, while they cover the field of view nearly homogeneously at long time.
At these time-lags, the differential frames at the two considered volume fractions look similar,
despite the huge difference in absolute time. 
More in detail, at $\Phi=0.79$, some degree of spatial correlations is already manifested at short time,
whereas it seems to be definitely negligible at $\Phi=0.65$,
in agreement with $\chi_4(10^{-1}\tau)$ being quite larger in the first case than in the second one (see Fig.~\ref{fig:fig1}h).
This difference, albeit minor, can be ascribed
to the fact that we are scaling time with respect to $\tau$, the characteristic time of $\alpha$-relaxation;
on the other hand, at early times, the dynamics is ruled by the $\beta$-relaxation, 
which is known to increase much less than $\tau$ on approaching the glass tranisition~\cite{Debenedetti,Cavagna}. 
In other words, the two frames at the lowest scaled time correspond
to different stages of the $\beta$-relaxation.
Also considering the long-time frames, some differences seem to be present in the spot "orientational order", loosely speaking.
 Some ordering at the largest volume fraction might be due to the tendency of the system to form
 short range crystal-like or "locally favoured" structures~\cite{Royall}.
Much clearer differences emerge between the differential frames at $\Delta t\simeq \tau$.
Indeed, in the central frame at $\Phi=0.79$,  spots corresponding to fast particles form large clusters that coexist
with $[\Delta I]^2\simeq0$-regions, where the system is still frozen,
suggesting that a liquid and a solid phase dynamically coexist on the timescale $\Delta t\simeq \tau$~\cite{Jstat16}.
At $\Phi=0.65$, instead, spatial correlations among spots remain rather small at $\Delta t\simeq \tau$,
according to the fact that $\chi_4$ is quite flat, with its maximum 
being an order of magnitude smaller as compared to the the maximum at $\Phi=0.79$.
Comparison between the central frames also suggests a "string-to-compact" crossover of the shape of dynamic clusters on approaching the glass transition~\cite{Woolines,gokhale2016localized}. 

\subsection{Comparison between non-Gaussian parameter and dynamic susceptibility}
\label{Subsec:ngp}

As extensively discussed in~\cite{DVA}, reliable measurements of $\chi_4$ via particle tracking are challenging, 
since the evaluation of the  mean square of a given dynamic correlation function, $<\Psi(t,\Delta t)^2>_t$, requires to track a large number of particles
simultaneously and for times much longer than the relaxation time. 
Even in ideal conditions, i.e., when particles are clearly resolved and video duration is long enough, as in our experiments,
trajectory ensembles with these features are difficulty recorded due, for example, to particles exiting the field of view and to
(even sporadic) failures of the tracking algorithm.
Thus, in order to spot out dynamic heterogeneities from particle tracking, alternative and easier to compute indicators should be exploited. 
These quantities, although less direct than $\chi_4$, 
can be computed by averaging over all recorded trajectories, no matter whether they are temporally overlapped or not.
Here we will use particle tracking to compute the lowest order non-Gaussian parameter, $\alpha_2$, that in 2D reads:

\begin{equation}
\label{eq:alpha_2}
\alpha_2(\Delta t)= \frac{\< r^4(\Delta t) \>}{2\<r^2(\Delta t)\>^2} - 1
\end{equation}
where $\< ~\>$ indicates average over particles and/or initial times.
The rationale behind this choice is as follows. 
In the presence of dynamic heterogeneities, the Van Hove function, 
i.e. the particle displacement distribution, differs from
the Gaussian distribution with zero mean and a variance equal to the mean square
displacement, which characterize standard Brownian diffusion. 
Specifically, the coexistence of long-lived fast and slow domains, characteristic of dynamic heterogeneities,
leads to fat tails in the particle distribution.
$\alpha_2(\Delta t)$ is the simplest way to characterize this deviation and its
temporal evolution, as it is defined through the ratio of first two moments of the distribution.
It should be emphasized that the non-Gaussian parameter, being based on the single particle
distribution function, cannot provide any \textit{direct} information on spatial correlations\textgx{:
indeed, to the best of our knowledge, no direct relation exists between the numerical value of $\alpha_2$ and the typical size of dynamic clusters.}
On the other hand, single-particle trajectory must somehow reflect dynamical correlations,
as it is known to be strongly intermittent  in glassy matter. 
Thus, $\alpha_2$ quantifies dynamic heterogeneities in this \textit{indirect} way.

\begin{figure}[ht]
\centering
\includegraphics[scale=0.3]{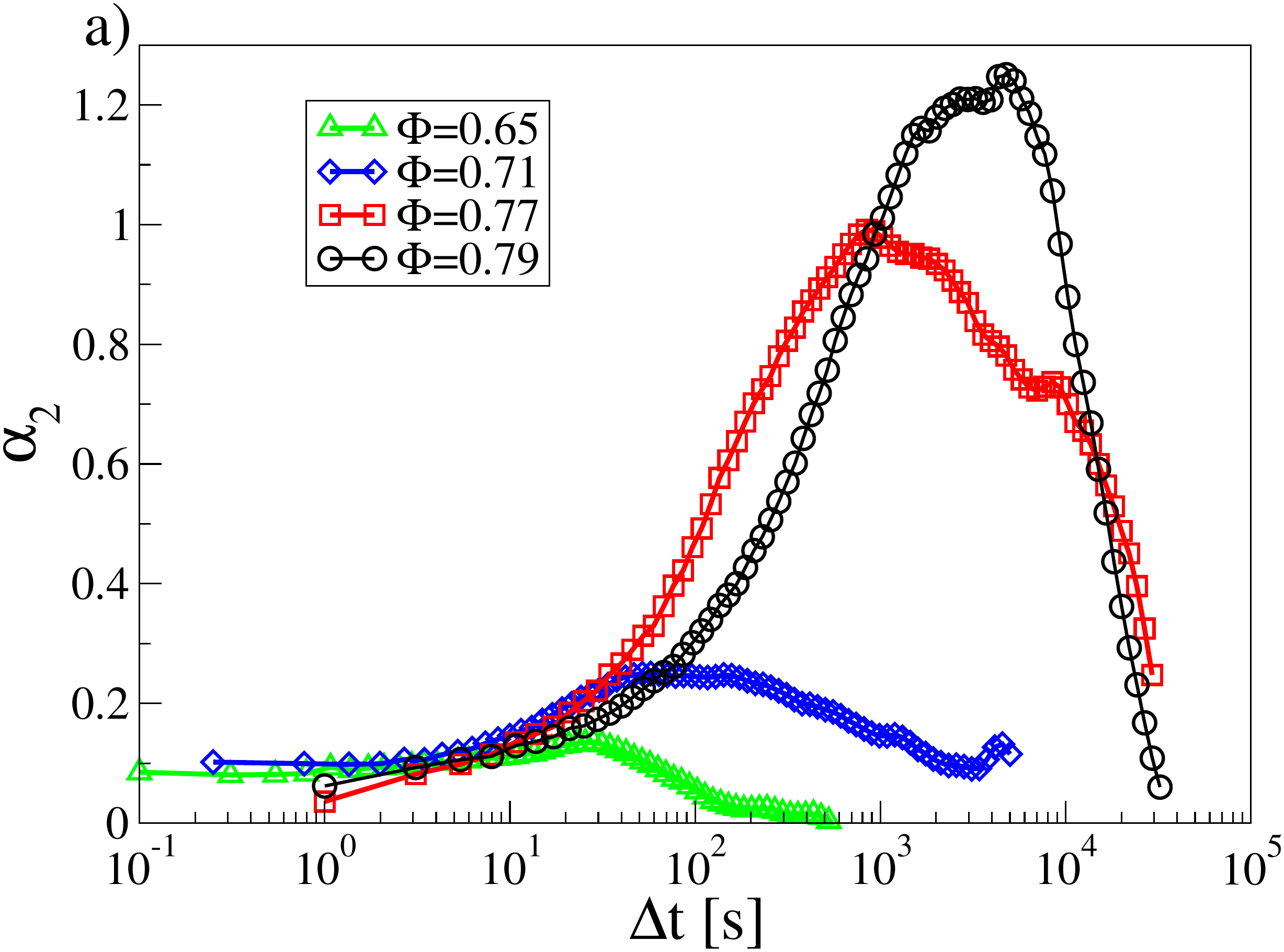}
\includegraphics[scale=0.3]{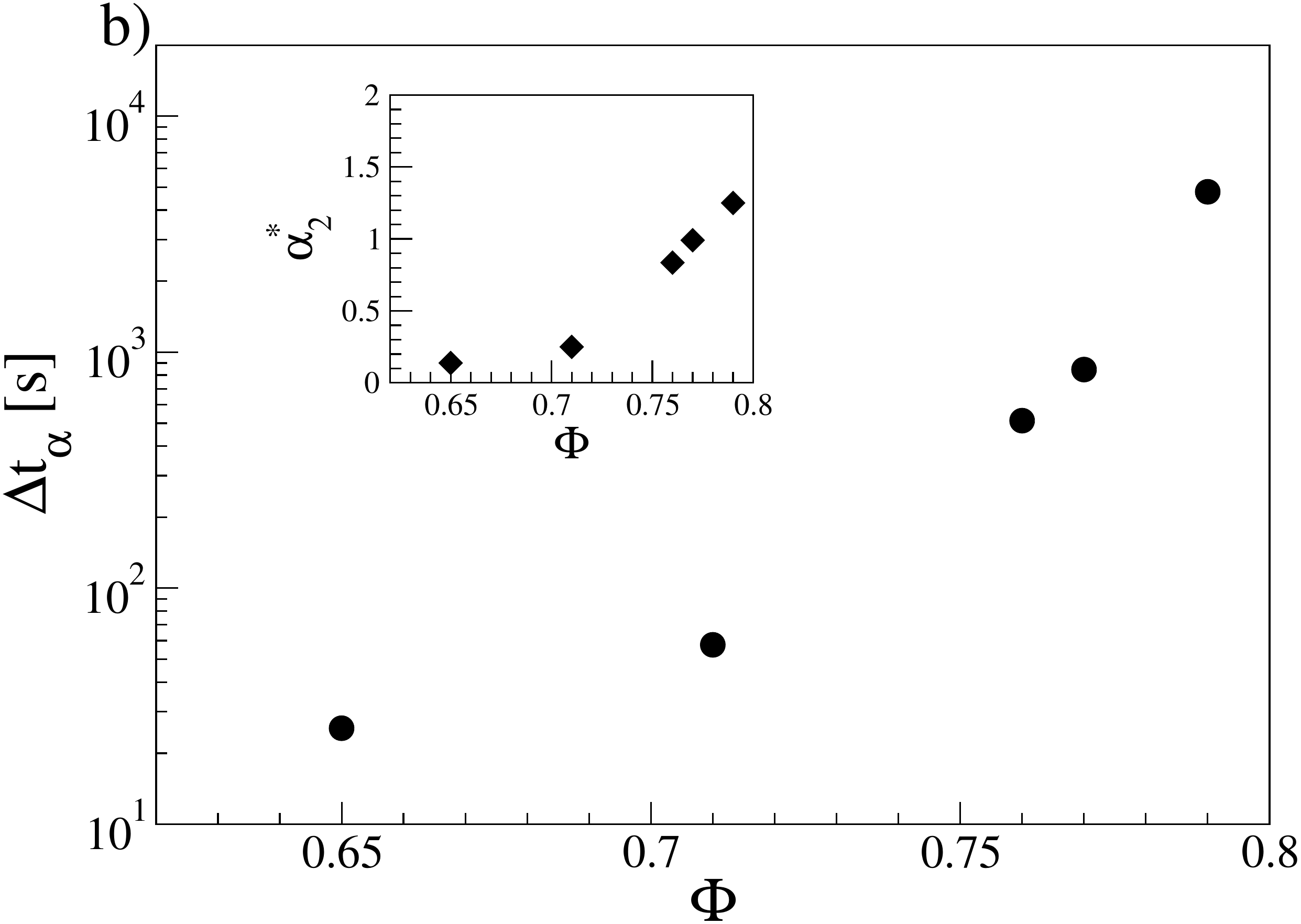}
\caption{\label{fig:ngp}
(a)Non-Gaussian parameter, $\alpha_2$, as function of the time-lag $\Delta t$ for different volume fractions, as indicated.
$\alpha_2(\Delta t)$ shows a maximum $\alpha_2 ^*$ at a time $\Delta t_{\alpha}$. 
(b) $\Delta t_{\alpha}$ as a function of $\Phi$.  
Inset: $\alpha_2 ^*$,  as a function of the volume fraction, $\Phi$.  
}
\end{figure}

Figure ~\ref{fig:ngp}a shows $\alpha_2$ as a function of $\Delta t$, for different volume fractions.
$\alpha_2(\Delta t)$ is nearly flat at small volume fraction,
to display a clear maximum at larger and larger times with increasing $\Phi$. 
This behaviour is readily understood, by considering
that single particle motion consists of an intermittent cage-jump dynamics.
The time shifting of the maximum corresponds to the average cage duration becoming 
longer and longer on approaching the glass transition; 
the growing of the maximum with $\Phi$ is due to the distribution of cage duration becoming broader and broader~\cite{SM15_exp,Jstat16,Baschnagel2014,Baschnagel2018},
leading to fatter tails of the Van Hove function.

The maxima of $\alpha_2(\Delta t)$, indicated by $\alpha_2^*$, and the corresponding times, $\Delta t_{\alpha}$,
are reported in Fig.~\ref{fig:ngp}b. 
$\Delta t_{\alpha}$ increases by three decades, while $\alpha_2^*$ grows over one decade \textgx{and is slightly larger than unity at the largest volume fraction.
Incidentally, we notice that $\alpha_2^*\simeq 1$ is consistent with the values reported 
for other glassy colloidal suspensions~\cite{WeeksScience,Chi_Gnan}}. 
Overall, the similarity between Fig.~\ref{fig:ngp}a,b and Fig.~\ref{fig:fig1}h,i is apparent:
qualitatively, at least, the two dynamic heterogeneity indicators $\alpha_2$ and $\chi_4$ share common features.

\begin{figure}[ht]
\centering
\includegraphics[scale=0.3]{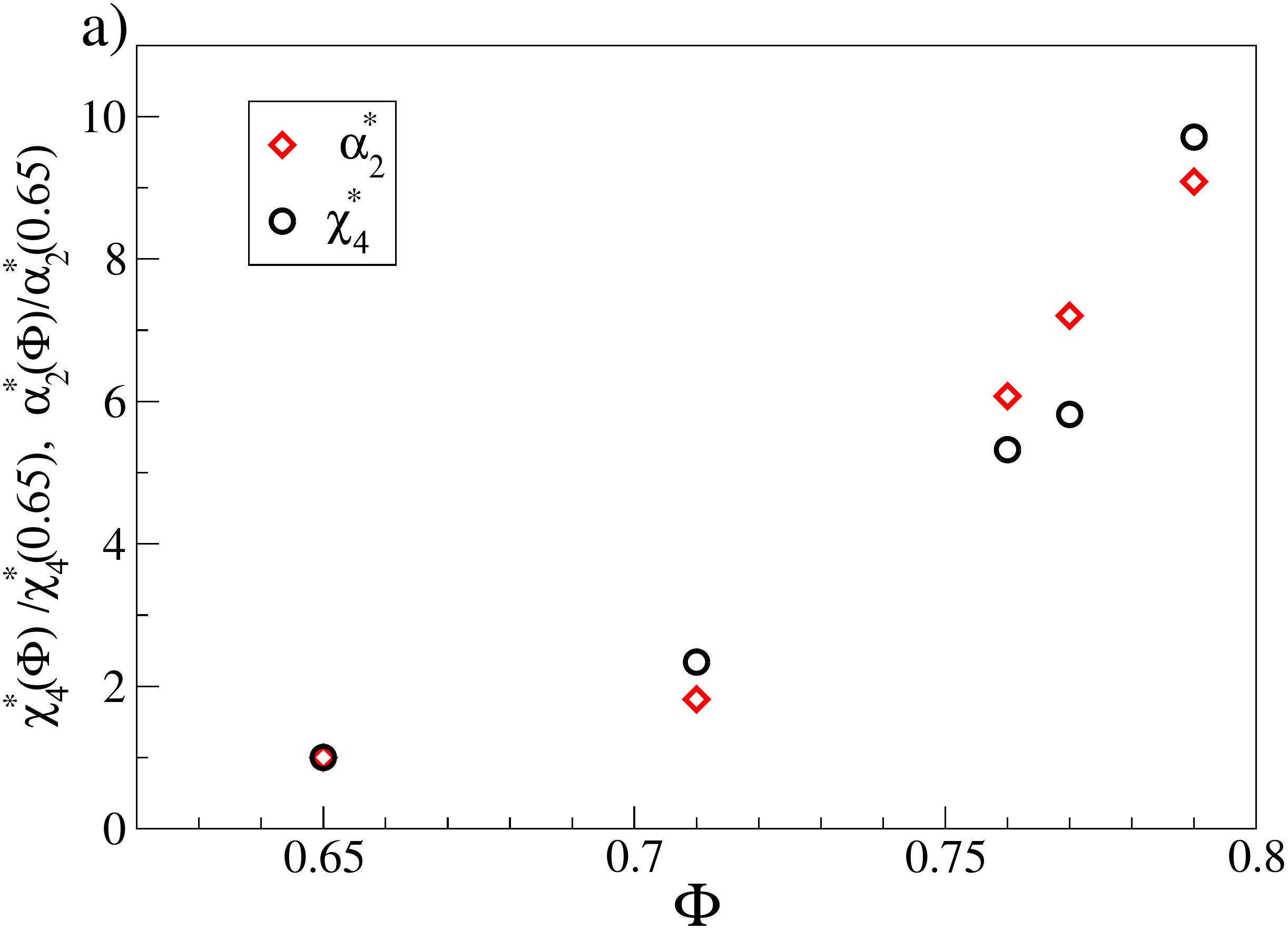}
\includegraphics[scale=0.3]{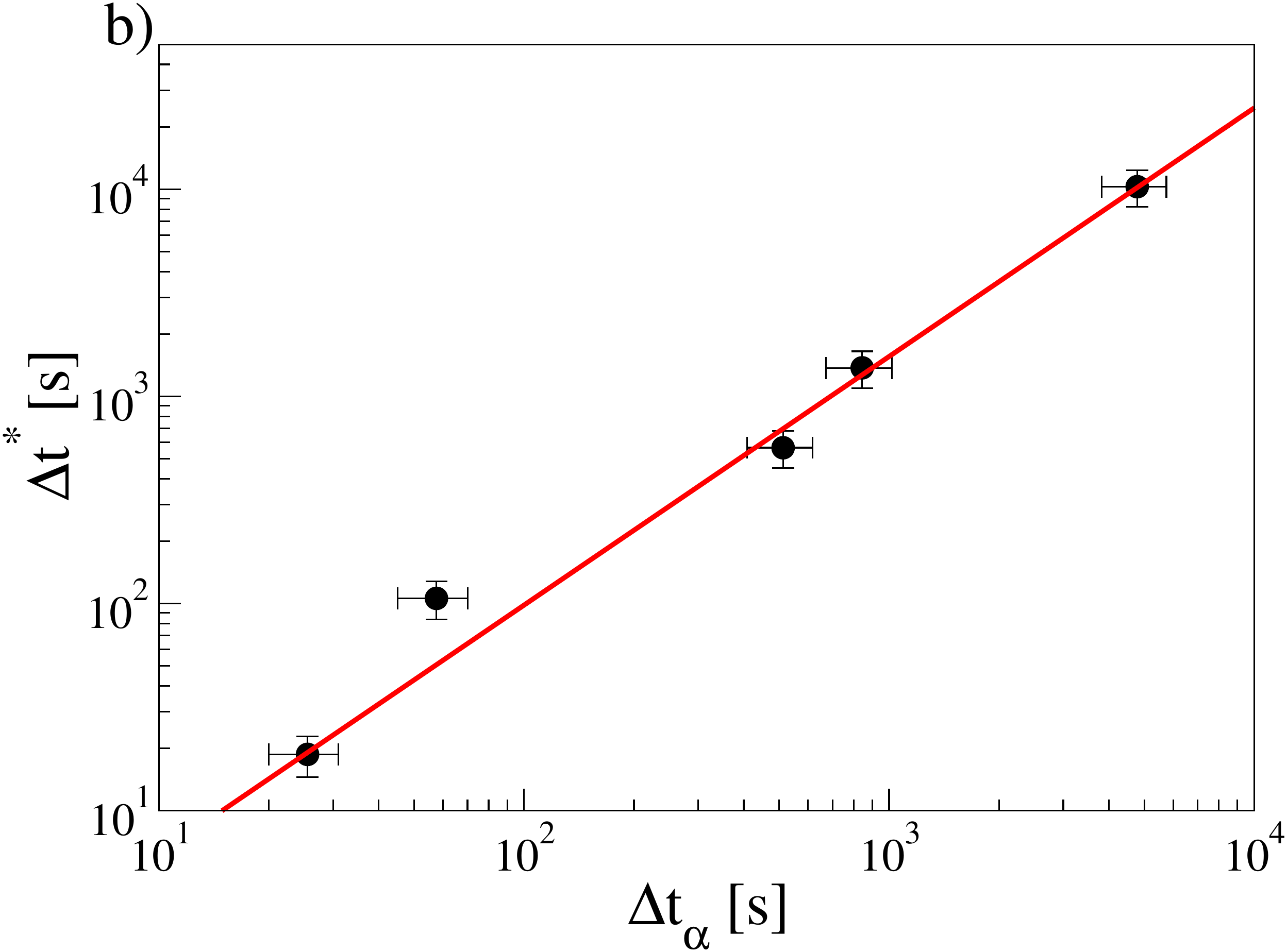}

\caption{\label{fig:chi_ngp}
(a) $\chi_4^*(\Phi)$ and $\alpha_2 ^*(\Phi)$ rescaled for their value at the lowest volume fraction in the plot, $\Phi=0.65$.
(b) Scatter plot of $\Delta t_{\alpha}(\Phi)$ vs $\Delta t^*(\Phi)$. 
The solid line is a fit $\Delta t^*=0.4\Delta t_{\alpha}^a$, with $a=1.20 \pm0.04$.
}
\end{figure}

To quantitatively check this similarity, we plot in Fig.~\ref{fig:chi_ngp}a both the maxima 
$\chi_4^*$ and $\alpha_2^*$ as a function of $\Phi$.  
Since $\chi_4^*$ is always larger than $\alpha_2^*$ by roughly an order of magnitude,
we rescale each of the two data sets with its own lowest volume fraction value. 
With this rescaling, the two data set nearly collapse on each other.
Fig.~\ref{fig:chi_ngp}b is a scatter plot of $\Delta t^*$ vs $\Delta t_{\alpha}$.
We choose this kind of plot here because both times span three orders of magnitudes with increasing volume fraction,
and therefore a power-law dependence, if any, would easily emerge. 
This is indeed the case, also unveiling that $\Delta t^*$ grows faster than $\Delta t_{\alpha}$ (even if not dramatically so),
because the slope in panel b is $1.2$.
Since we have already demonstrated in Fig.~\ref{fig:fig1}i that $\Delta t^* \simeq \tau \propto (\Phi_c -\Phi)^{-2.5}$,
we conclude that also $\Delta t_{\alpha}$ follows a Mode Coupling Theory behaviour, although with a smaller exponent $\simeq -2.1$.

\begin{figure}[ht]
\centering
\includegraphics[scale=0.3]{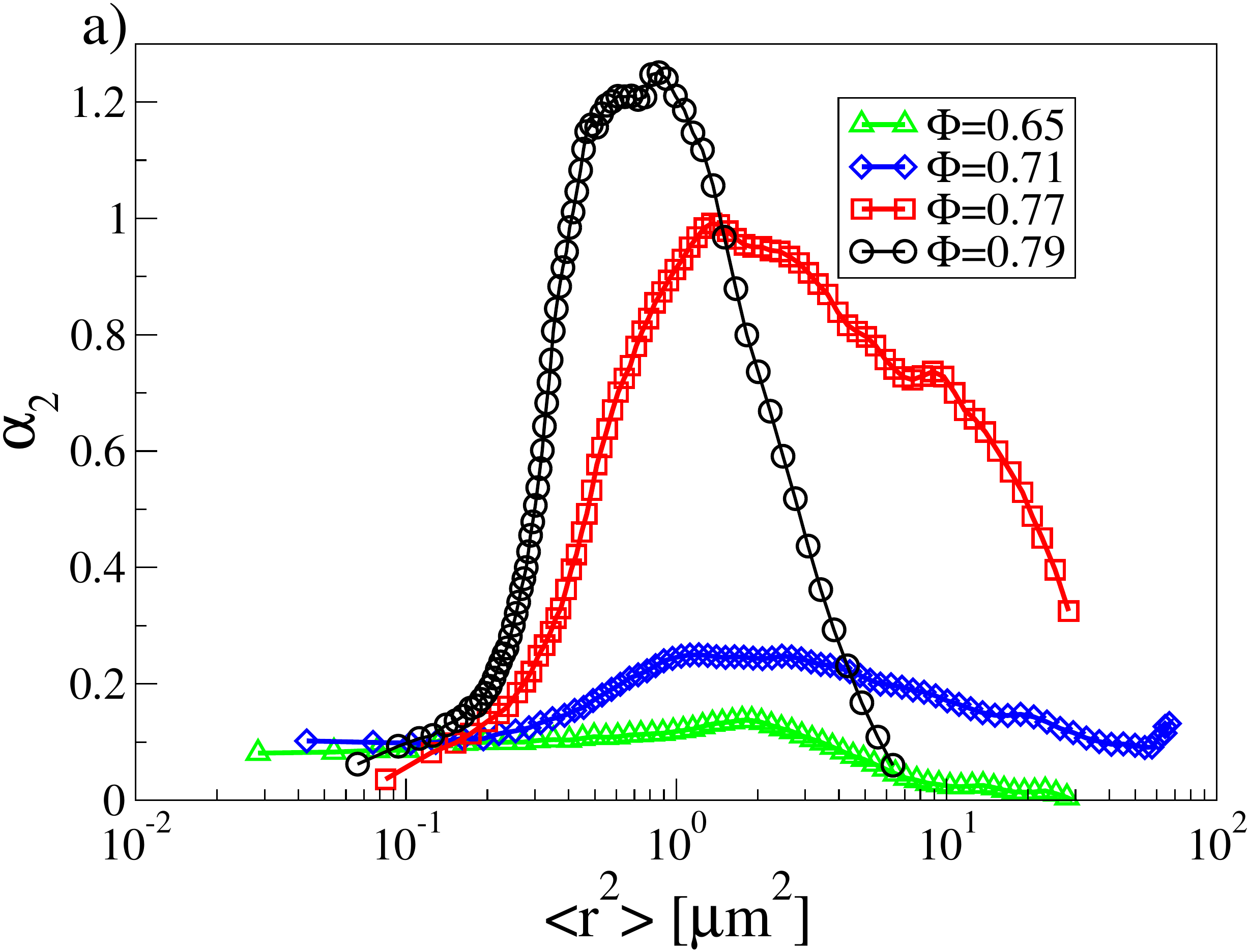}
\includegraphics[scale=0.3]{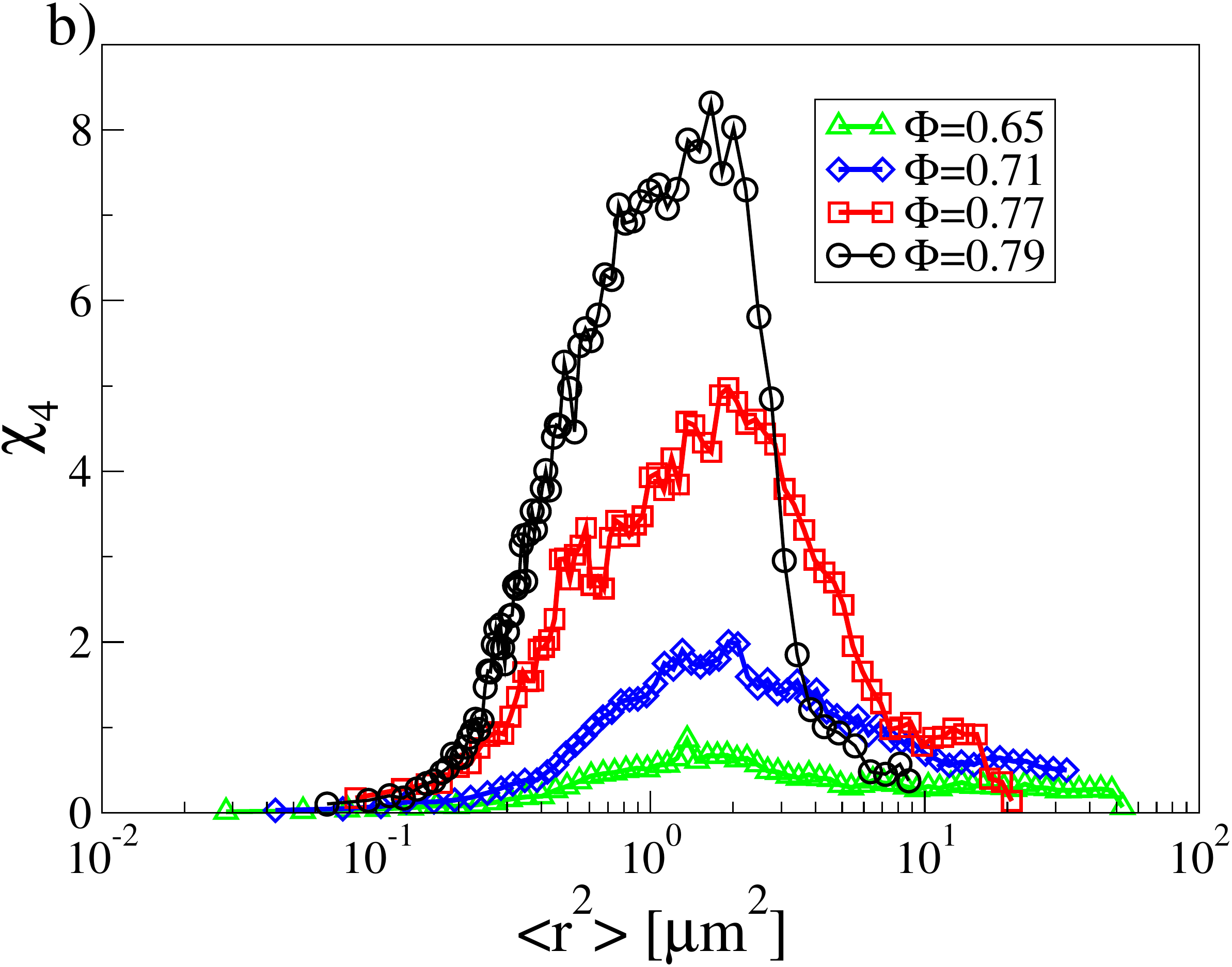}
\caption{\label{fig:DH_msd}
Scatter plot of (a) the non-Gaussian parameter, $\alpha_2(\Delta t)$, and (b) the dynamic susceptibility, $\chi_4(\Delta t)$,
vs  the mean square displacement, $\<r^2(\Delta t)\>$, for different volume fractions, as indicated.
}
\end{figure}

Since $\alpha_2$ is related to the single particle cage-jump motion,
a rough estimate of the typical cage size can be obtained by plotting  $\alpha_2(\Delta t)$ as a function of $\<r^2 (\Delta t)\>$ 
and  \textgx{considering the mean square displacement, $\<r^2 (\Delta t_{\alpha})\>$, 
which corresponds to the maximum of the non-Gaussian parameter}~\cite{Chi_Gnan}.
Fig.~\ref{fig:DH_msd}a shows this kind of plot for our system:
We do observe that a backgoing shift \textgx{of the position of the maxima (on the x-axis)} occurs on increasing the volume fraction.
Estimating the cage size (area) in this way, we infer that cage-shrinking is taking place, by about a factor 3.
It should be immediately noticed that such a factor of shrinking is much larger than the reduction in the square interparticle distance (just a few percent)
ensuing from the change in volume fraction.
Indeed, the cage size estimated from panel a is not a static quantity, but a dynamical one. 
The present result is fully consistent with a previous detailed analysis on this same system, based on segmentation of
trajectories in cage and jumps~\cite{SM15_exp}.
On the other hand, $\alpha_2$-based cage-shrinking factors significantly larger than our own were estimated for a glassy suspension 
of soft particles, where the system can be compressed above jamming~\cite{Chi_Gnan}. 

It is now tempting to examine the behaviour of $\chi_4$ as a function of $\<r^2\>$; 
this is shown in Fig. ~\ref{fig:DH_msd}b. Here, we do not observe any shift of $\chi_4^*$
on changing the volume fraction. 
This can be rationalized by considering that the dynamic susceptibility 
is determined not only by single particle cage-jump motion, but also by the evolving
of correlation between different particle displacements.
The lack of variation in the position of the maximum at different volume fractions is consistent
with the fact that our experiments are essentially in the Mode Coupling regime.
Accordingly, no clear deviation from Stokes-Einstein relation is found~\cite{colsua17}: $\tau(\Phi) \propto D^{-1}(\Phi)$.
Since we find $\Delta t^* \simeq \tau$, it follows that 
$\<r^2(\Delta t^*)\> \propto D \Delta t^* \propto D D^{-1}=const$.

\section{Discussion and Conclusion}
\label{sec:discussion}

For systems in which primary particles can be clearly identified, the probing length of DVA
 is  the particle size and, indeed, the outcome of DVA are very similar to those expected from Dynamic Light Scattering, 
(or similar wavelength tunable tecniques) at a wavelength $\lambda$ of the order of the particle size.
This does not imply that the probing length is the only relevant length scale in general and, indeed, 
 other length scales may affect the outcome of DVA. 
 For systems in which particles are caged by their neighbours or by other physical constraints (e.g. tracers diffusing in a porous materials), 
another relevant length scale surely is the amplitude of the oscillation within the cage, 
In this case, a two step decay is generally expected for the dynamical order parameter $Q(\Delta t)$,
with the first decay ($\beta$ relaxation) being related to the particle rattling within the cage, and the final decay ($\alpha$ relaxation) due to displacements out of the cage. 
For {\it crowded} repulsive systems, however, such as our hard-sphere suspension, the interparticle distance 
is just slightly larger than the particle size (i.e. particles are essentially in contact), 
and, therefore, the oscillation within the cage is very small. Accordingly, the Dynamic order parameter is poorly sensitive to rattling 
(the two step decay is indeed not clearly manifested),
and its decay appears to be controlled by a single characteristic length, i.e. the particle size or, equally, the interparticle distance.

In this paper, we described how to use DVA  to quantitatively characterize and 
directly visualize the relaxation dynamics of glassy suspensions of Brownian hard spheres in water. 
Starting from the variance of the differential frames we introduced a 
dynamic order parameter, $Q(\Delta t)$, to describe the relaxation process, and
measured the structural relaxation time, $\tau$, as well as the dynamic susceptibility $\chi_4(\Delta t)$.
The DVA method also leads to directly visualize the relaxation process and the emergence of dynamic heterogeneities,
by focusing on patterns in the differential frames. 
Such patterns essentially correspond to clusters of fast particles, 
which had previously been identified through more complicated approaches~\cite{WeeksScience}.
It is worth remarking that, to effectively visualize dynamic heterogeneities, it is crucial to choose a well defined timescale, namely, $\Delta t^*$, corresponding to the
maximum in the dynamic susceptibility, $\chi_4^*$.
Such a timescale is self-determined by the dynamics and, therefore, will change on varying the system control parameter ($\Phi$ in our experiment).
As discussed above, $\chi_4^*$ and $\Delta t^*$ are rough estimates of the typical size and life-time, respectively, of a cluster of dynamically correlated particles.
Indeed, $\chi_4(\Delta t)$ is also defined as the space integral of a correlation function, 
$G_4(r, \Delta t)$, measuring correlations of the displacements at $\Delta t$,
between particles separated by a distance $r$~\cite{DHbook}.

In this work, we have also measured another popular indicator of dynamic heterogeneities, the non-Gaussian parameter $\alpha_2$, through particle tracking.
$\chi_4$ and $\alpha_2$ are usually assumed to give similar information on dynamic heterogeneities.
Here, by direct comparison, we have shown that those two quantities in fact display 
 some quantitative differences.
This can be understood in terms of the different nature of the two indicators.
$\alpha_2$ is indeed closely related to single-particle cage-jump dynamics, 
but cannot provide direct information on spatial correlations~\cite{SM15_spcorr}, as $\chi_4$ does.
Particularly revealing is, in our opinion, the difference found when plotting these two quantities as
a function of the mean square displacement: when increasing the volume fraction, 
the maximum of  $\alpha_2$ shifts to lower $\<r^2\>$, indicative of cage-shrinking, whereas $\chi_4^*$ does not.
It would be interesting to check if this difference is maintained in compressed soft particle system,
in the presence of stronger cage-shrinking in more glassy states.

The system here investigated is the simplest but, hopefully, paradigmatic model system for glassy dynamics.
Indeed, the scenario just illustrated for $\chi_4$ is qualitatively found in a wide variety of liquids, even including
supercoleed water~\cite{perakis2017diffusive}.  In addition, we notice that experimental and numerical studies 
have very recently highlighted the relevance of cage-jump dynamics, at the atomic scale,  
for supercooled water~\cite{perakis2018coherent,kikutsuji2018hydrogen,kikutsuji2019diffusion}.

As a future perspective, DVA can be easily exploited to investigate a variety of soft materials,
such as colloidal gels, emulsions and foams, as well as red blood or epithelial cells.
Of course, it will be especially useful for systems where single particles cannot be resolved:
we are currently investigating a dense suspension of sub-micron charged vesicles.
Finally,  we suggest that combining  DVA and Differential Dynamic Microscopy  could allow dynamic fluctuations in a range of length-scale to be measured.


\section{Appendix A1. General video requirements}
\label{sec:video_req}
The approach proposed in this paper can be applied  to soft matter samples at equilibrium or in steady state conditions, or to sample that can be
considered quasi-stationary, over the timescale necessary to properly measure $Q(\Delta t)$ and $\chi_4(\Delta t)$. 
This is the case, for example, of coarsening foams, like those investigated in Ref.~\cite{CipellettiPRL04}. 

\begin{figure}[ht]
\centering
\includegraphics[scale=0.5]{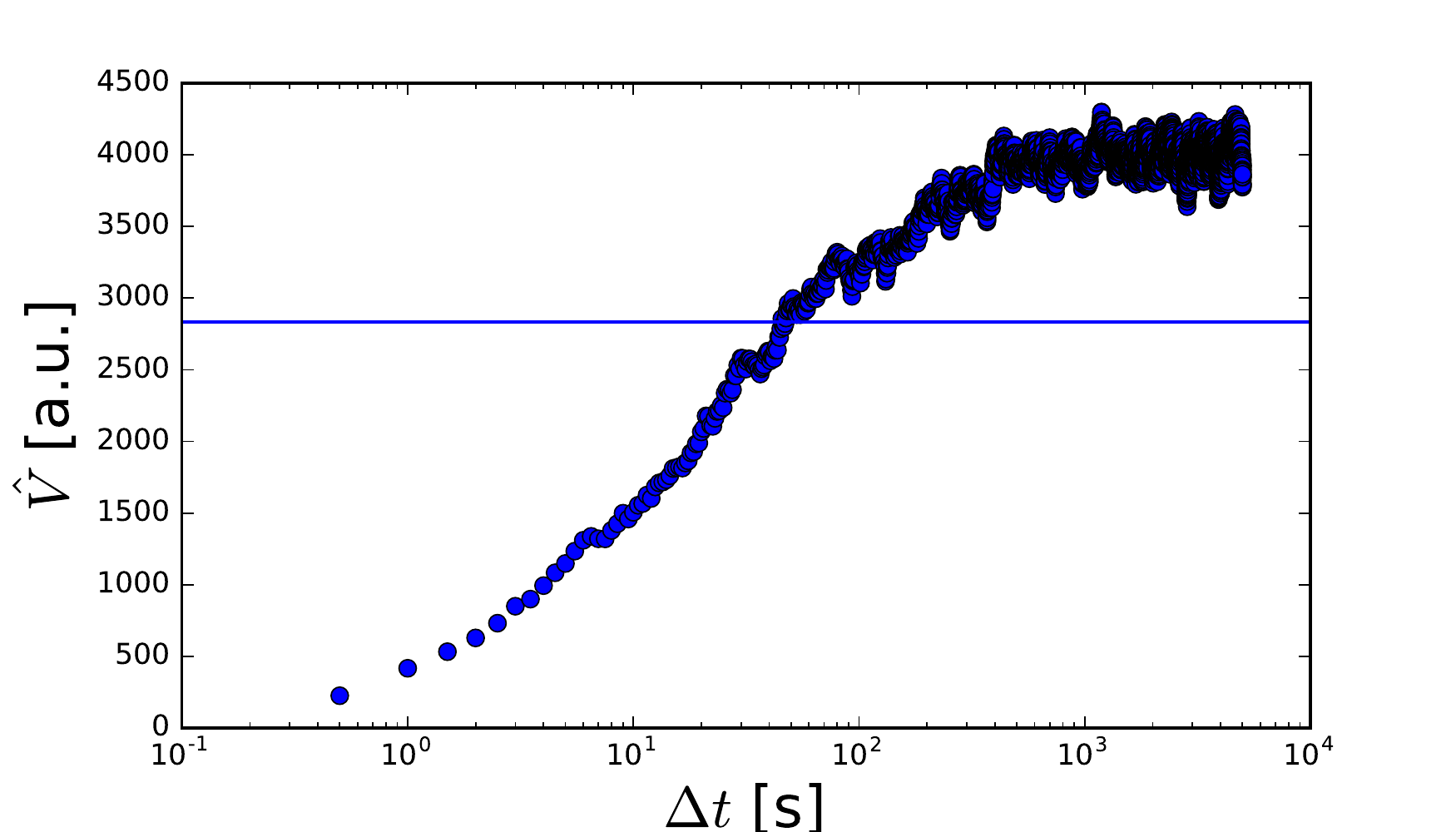}
\caption{\label{fig:unav_v}
Temporally non-averaged variance $\hat V(0, \Delta t)$ for the same systems considered in Fig.~\ref{fig:fig1}f. 
A preliminary estimation of the relaxation time can be obtained
from the intercept of the solid line, corresponding to $2/3$ of the maximum variance,   and the data points.
}
\end{figure}

In this appendix, we illustrate some properties a video should have to be suited to DVA.
We indicate with $\hat \tau$ an estimate, even a rough one, of the structural relaxation time of the sample under consideration,
with $t_v$ the overall video duration, and with $t_f$ the time interval between subsequent frames.  

 \textgx{As regards the temporal features}, a video should comply with the following requirements for DVA to be fully applicable: 
 {\it i)} $t_v>\hat \tau$   (e.g. $t_v=10\hat \tau$ or larger);  {\it ii)} $t_f\ll \hat \tau$ (e.g. $t_f=0.01\tau_{br}$ or smaller).

Condition {\it i)} is needed to perform appropriate temporal averages, namely averaging over a number of configurations that are separated by time of the order of the relaxation time and can be therefore considered as uncorrelated.
Condition {\it ii)}  is desired to have a detailed monitoring of the relaxation process temporal evolution. 
\textgx{Indeed, the value of $t_f$ also sets the temporal resolution.}

To check whether  a video complies with the above criteria, a broad estimation of the structural relaxation time, at least as order of magnitude, is needed. 
In lack of expectations based on previous experiments, a DVA inspired method can be used to have a  preliminary estimate of the relaxation time. 
This consists in calculating the non-averaged (over time) differential variance, $\hat V(0, \Delta t)$, and in estimating $\hat \tau$ as the time where the variance roughly attains $2/3$ of its maximum value.
This is illustrated in Fig.~\ref{fig:unav_v}, which refers to the same systems of Fig.~\ref{fig:fig1}f.
If the variance does not attain a plateau, the considered video is probably too short, i.e.  the video does not complies with {\it (i)}. 
Conversely, If the variance appears flat, the acquisition rate is probably too large,  i.e.  the video does not complies with {\it (ii)}. 

\textgx{Concerning the spatial video requirements, in order 
 to obtain a ÒfairÓ monitoring of the structural relaxation (i.e., sensitive to primary particles rearrangements),
the pixel size should be of the same order or smaller than the primary particle size.
 Under this condition, indeed, one gets a clear-cut signal if a single particle leaves a pixel.
In the investigated video, for example, the ratio between pixel size and average particle radius is slightly larger than $0.1$. 
In case of worse resolution, DVA can still be sensitive to the relaxation of dynamic clusters, if any, equal to or larger than the pixel size.}

\section{Appendix A2. DVA and ACII}
In this section, we discuss the relation between DVA and ACII~\cite{Durian_SM, Durian_PRE}.
In particular, we clarify under what conditions and approximations the DVA and the ACII dynamic order parameter coincide, and
how some differences emerge otherwise.

The Dynamic order parameter of ACII is defined as:
\begin{equation}
\label{eq:H}
\hat H(t, \Delta t)=\frac { \< I({\bf r},t+\Delta t) I({\bf r},t)\>_r - \< I({\bf r},t)\>_r^2 } {\< I({\bf r},t)^2 \>_r - \< I({\bf r},t)\>_r^2 }
\end{equation}
where $\< ~\>_r$ indicates average in space.

The Dynamic order parameter of DVA has been given in Eq.~\ref{eq:overlap}.
It is :
\begin{equation}
\hat V(t, \Delta t) =\frac{1}{L^2} \int_{L^2} dr^2 \Delta I^2({\bf r},t,\Delta t)=\< \Delta I^2({\bf r},t,\Delta t) \>_r
\end{equation}

Performing some calculations on this average, 
\begin{equation}
\begin{split}
\< \Delta I^2({\bf r},t,\Delta t) \>_r = \< \left [ I({\bf r},t+\Delta t)- I({\bf r},t) \right ]^2 \>_r 
&=\< I({\bf r},t+\Delta t)^2 \>_r + \< I({\bf r},t)^2 \>_r - 2\< I({\bf r},t+\Delta t) I({\bf r},t)\>_r, 
\end{split}
\end{equation}

and assuming that in stationary conditions $\< I({\bf r},t+\Delta t)^2 \>_r = \< I({\bf r},t)^2 \>_r$, we find:
\begin {equation}
\hat V(t, \Delta t) = 2 \left[ \< I({\bf r},t)^2 \>_r - \< I({\bf r},t+\Delta t) I({\bf r},t)\>_r  \right ]
\end{equation}

For ergodic or liquid-like systems, such as those investigated in our experiment,
the long-time plateau value of the temporally averaged variance, $V_{\infty}=\lim_{\Delta t \rightarrow \infty} V(\Delta t)$, reads:
\begin {equation}
V_{\infty} =  \left \< \lim_{\Delta t \rightarrow \infty} \hat V(t, \Delta t) \right \>_t=  \left \< 2 \left[ \< I({\bf r},t)^2 \>_r - \< I({\bf r},t)\>_r^2  \right ] \right \>_t
\end{equation}
since ergodicity guarantees that
\begin{equation}
\label{eq:erg} 
 \lim_{\Delta t \rightarrow \infty} \< I({\bf r},t+\Delta t) I({\bf r},t)\>_r =\< I({\bf r},t+\Delta t)\>_r \< I({\bf r},t)\>_r= \< I({\bf r},t)\>_r^2 
 \end{equation}
 
This provides a way to calculate the quantity, $V_{\infty}$, involved in the definition of the dynamic order parameter, without directly probing the long time plateau of the variance.
Incidentally, we note that the term between $\<\>_t$ of the rhs of the last equation does not depend on $\Delta t$ and, therefore, can be considered a structural quantity, not a dynamic one.
In other words, $\left[ \< I({\bf r},t)^2 \>_r - \< I({\bf r},t)\>_r^2  \right ]$ only depends on the system configuration at time $t$, not on the difference between configurations
evaluated at two different times. For equilibrium systems, like those investigated in this paper, the structure does not change in time, all explored configurations being statistically equivalent.
Accordingly,  $\left[ \< I({\bf r},t)^2 \>_r - \< I({\bf r},t)\>_r^2  \right ]$ may be expected to poorly fluctuate in time and equivalence 
with the corresponding temporally averaged quantity assumed as an approximation,
$\< 2 \left[ \< I({\bf r},t)^2 \>_r - \< I({\bf r},t)\>_r^2  \right ] =   \left \< 2 \left[ \< I({\bf r},t)^2 \>_r - \< I({\bf r},t)\>_r^2  \right ] \right \>_t$.
Using this approximation in eq.~\ref{eq:overlap}, $\hat Q(t, \Delta t)$ reads:
\begin{equation}
\label{eq:dva_acii}
\begin{split}
\hat Q(t, \Delta t)&=1-\frac{\hat V(t, \Delta t)}{V_{\infty}}  
=1- \frac{ \< I({\bf r},t)^2 \> - \< I({\bf r},t+\Delta t) I({\bf r},t)\>} {\< I({\bf r},t)^2 \> - \< I({\bf r},t)\>^2 }
= \frac { \< I({\bf r},t+\Delta t) I({\bf r},t)\> - \< I({\bf r},t)\>^2 } {\< I({\bf r},t)^2 \> - \< I({\bf r},t)\>^2 },
\end{split}
\end{equation}
that, to the consider approximation, formally equals the Dynamic order parameter of ACII (Eq.~\ref{eq:H}).

\begin{figure}[ht]
\centering
\includegraphics[scale=0.33]{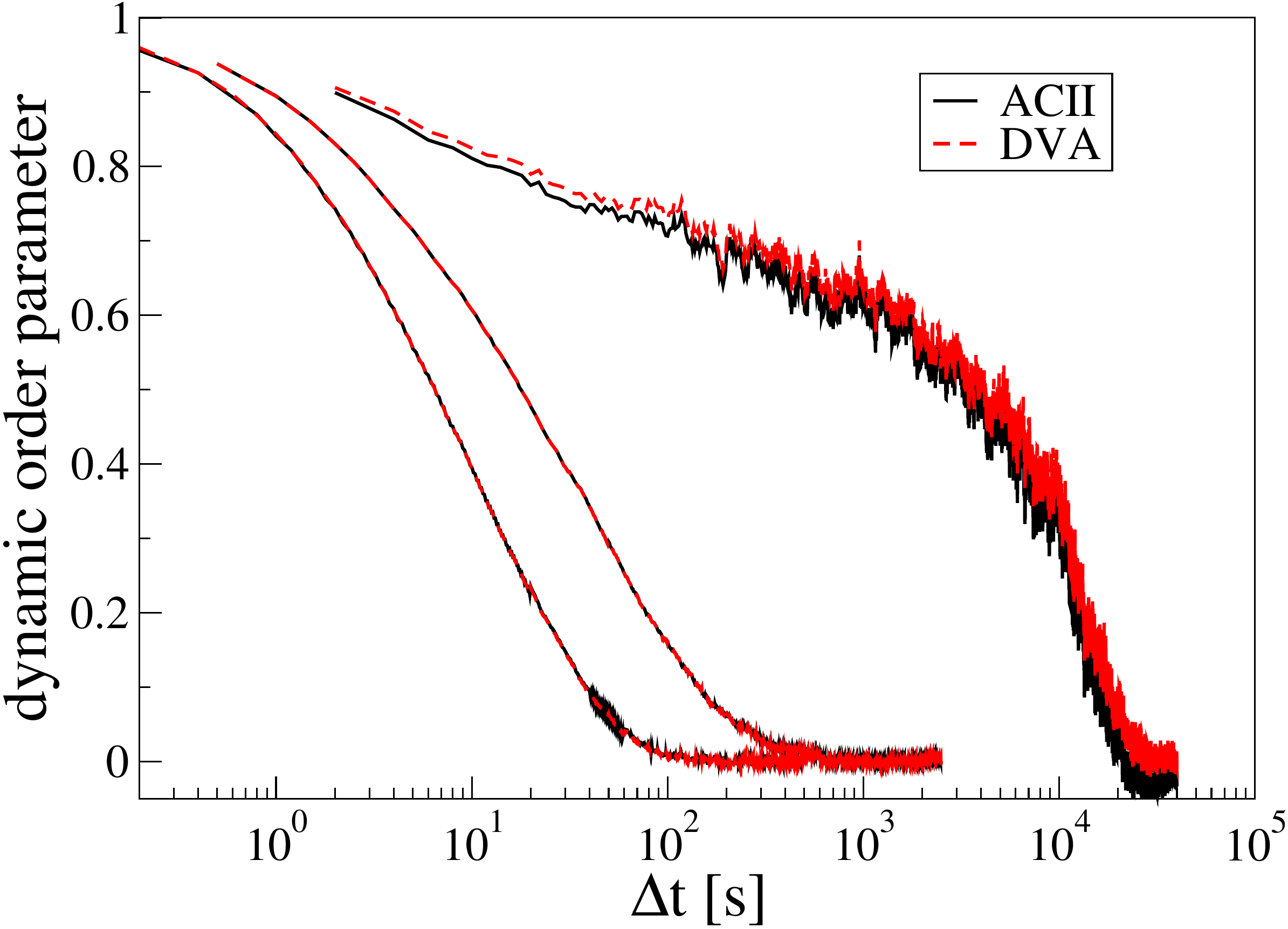}
\caption{\label{fig:comp}
ACII (solid line) and DVA (dashed line) dynamic order parameter as a function of $\Delta t$ and for volume fractions, $\phi=0.65$, $0.71$ and $0.79$, from left to right, respectively.
}
\end{figure}

Figure ~\ref{fig:comp} shows that the dynamic order parameters  as defined by ACII and DVA respectively, are very similar over the whole range of time-lag and volume fraction investigated.
This result demonstrates that the hypothesis done in the calculations above holds in very good approximation for our colloidal suspensions,
as expected from the fact that such a system is liquid-like and at equilibrium, with the dynamics being time translationally invariant.
Incidentally, it is worth remarking that the equality of Eq.~\ref{eq:dva_acii} is based on Eq.~\ref{eq:erg} and, therefore, on the hypothesis that the investigated system
can be considered fully ergodic on the considered observation time.
If this condition is not fulfilled, DVA and ACII may lead to completely different results. 
For example, in the case of an ergodic component of particles diffusing within an immobile/non ergodic matrix~\cite{Granick_nano},
ACII is strongly affected by the presence of the matrix, whereas DVA enables to isolate and highlight the relaxation of the ergodic component, 
since image difference cancels out the contribution of the immobile matrix.
In particular, the ACII dynamic order parameter is expected to show a partial decay to a finite plateau, 
which is due to the contribution of the non-ergodic component.
Conversely, the DVA dynamic order parameter keeps describing the relaxation of the mobile component and fully decay to zero.
For practical purposes, the outcome of ACII may be affected significantly by the presence of dust particles (typically located
on the protective glass of the camera sensor and resulting in dark spots in the image ~\cite{DDM}),
since these dust particles will also result in a non-ergodic, but obviously artificial, contribution.
By contrast, performing image difference, as in DVA,  eliminate this artificial, static contribution.

In addition, It is worth noticing that the primary signals provided by image difference  is known to have better performances than ACII, 
since it is less sensitive to low-frequency noise and drifts in the intensity~\cite{Schatzel, Schatzel2, Lorusso, Sprakel}.
Finally, we are not aware of any paper presenting direct visualization of the relaxation process and of dynamic heterogeneities obtained by ACII.
 
\end{document}